# Improved Motor Imagery Classification Using Adaptive Spatial Filters Based on Particle Swarm Optimization Algorithm

Xiong Xiong[1], Ying Wang[1], Tianyuan Song[1], Jinguo Huang[1]*, Guixia Kang[1]*

1. School of Information and Communication Engineering, Beijing University of Posts and Telecommunications, Beijing 100876, China

   Corresponding author: Jinguo Huang and Guixia Kang

## Abstract

As a typical self-paced brain-computer interface (BCI) system, the motor imagery (MI) BCI has been widely applied in fields such as robot control, stroke rehabilitation, and assistance for patients with stroke or spinal cord injury. Many studies have focused on the traditional spatial filters obtained through the common spatial pattern (CSP) method. However, the CSP method can only obtain fixed spatial filters for specific input signals. Besides, CSP method only focuses on the variance difference of two types of electroencephalogram (EEG) signals, so the decoding ability of EEG signals is limited. To obtain more effective spatial filters for better extraction of spatial features that can improve classification to MI-EEG, this paper proposes an adaptive spatial filter solving method based on particle swarm optimization algorithm (PSO). A training and testing framework based on filter bank and spatial filters (FBCSP-ASP) is designed for MI EEG signal classification. Comparative experiments are conducted on two public datasets (2a and 2b) from BCI competition IV, which show the outstanding average recognition accuracy of FBCSP-ASP. The proposed method has achieved significant performance improvement on MI-BCI. The classification accuracy of the proposed method has reached 74.61% and 81.19% on datasets 2a and 2b, respectively. Compared with the baseline algorithm (FBCSP), the proposed algorithm improves 11.44% and 7.11% on two datasets respectively. Furthermore, the analysis based on mutual information, t-SNE and Shapley values further proves that ASP features have excellent decoding ability for MI-EEG signals, and explains the improvement of classification performance by the introduction of ASP features.

# 1 Introduction

Brain-computer interface (BCI) technology is an emerging field that allows direct connection between the brain and external devices [1, 2]. BCIs have many potential applications, including assisting paralyzed patients, operating machines in extreme environments, and controlling neuroprosthetic limbs [3-5]. Non-invasive electroencephalogram (EEG) signals break away from the ethical constraints and the requirements of invasive brain surgery, and become a more suitable way to construct BCI for normal people [6-9]. For non-invasive brain-computer interfaces, sensorimotor rhythms [10], event-related potentials [11], and steady-state visual evoked potentials [12] are the three main application paradigms. Motor imagery (MI) is a common method used by humans to evoke sensorimotor rhythms in an autonomous way [13, 14]. Motor imagery tasks inhibit contralateral sensorimotor areas of the brain. MI-based systems have shown great potential in helping patients with stroke [15, 16], spinal cord injuries [17-19], and amyotrophic lateral sclerosis [20, 21].

For MI-BCI, accurate decoding of user intentions is crucial for the practicability and robustness of BCI systems. How to extract effective features related to motor imagery in EEG signals is the key to accurate decoding [1, 7]. Some researchers have used event-related desynchronization and event-related synchronization (ERD/ERS) to classify mental states [22, 23]. However, the low signal-to-noise ratio (SNR) of EEG signals affects the detection of ERD/ERS patterns. In addition, due to the topology of motor neurons, the EEG signals collected from the cerebral cortex are usually mixed by multiple sensorimotor neurons, resulting in poor spatial resolution of the original EEG signals and reducing the pattern recognition performance [24].

To improve the spatial resolution of ERD suppression and ensure the performance of pattern recognition, commonly used feature extraction methods include spectral analysis [25], autoregressive [26], source reconstruction [27], and common spatial pattern (CSP) [28]. Among them, CSP features have been widely used in MI-BCI. The algorithm transforms the EEG signal by solving an optimal spatial filter to maximize the variance of one MI task and minimize the variance of the other MI task. Therefore, the CSP algorithm is suitable for feature extraction of multi-variable EEG signals [29].

However, the traditional CSP features have the problem of over-fitting, so some methods improve the effect of CSP algorithm by weighting or regularizing CSP features. Regularized Common Spatial Pattern (RCSP) improves classification accuracy by combining CSP with ridge regression and regularization [30]. Furthermore, FBRCSP introduces a filter bank based on RCSP, and uses feature selection based on mutual information to reduce the dimension, to improve the recognition effect of MI. Discriminative FBCSP (DFBCSP) achieves high classification performance by combining FBCSP with discriminative classifiers [31]. Sparsity FBCSP (SFBCSP) incorporates sparsity constraints into FBCSP to enhance feature selection and reduce feature space dimension [32]. Furthermore, Local Region Frequency CSP (LRFCSP) extracts features from specific frequency bands in local brain regions to improve classification accuracy [33]. Spectrally weighted CSP (SWCSP) weights the contribution of different frequency bands by considering the correlation of frequency features [34]. Penalized Time-frequency band CSP (PTFBCSP) is similar to SWCSP by penalizing irrelevant features to improve classification accuracy, but PTFBCSP further considers irrelevant features in time and frequency domains [35].

Besides, some studies consider both sequence relationships and frequency bands to enhance CSP. For example, the most representative research includes Separable CSSP (SCSSP), which improves classification accuracy by simultaneously considering sequence relationships and frequency bands in

EEG signals [36]. CSP based on the longest continuous repeated sliding window (LCR-SW-CSP) improves CSP features through multiple time windows, thereby enhancing the classification accuracy when processing MI-EEG [37]. Temporal Constrained Sparse Group Spatial Pattern (TSGSP) uses temporal constrained sparse constraints to extract spatial features [38]. Time-frequency CSP (TFCSP) extracts CSP features in time-frequency domain to obtain more effective features [39].

Furthermore, other studies enhance or extend CSP features to improve classification accuracy. The most commonly used method in this category to enhance CSP features is to extract spatial features while considering frequency bands. The most representative research is the filter-bank CSP (FBCSP) [40]. FBCSP obtains CSP features over different frequency bands by introducing filter banks into EEG signals before using CSP feature extraction. There are some other methods that also consider frequency domain information include Common Spatial-spectral Pattern (CSSP), which considers both spatial and spectral information in EEG signals [41]. CSSP aims to identify a set of spatial filters that can capture spatial and spectral features specific to a given task or class. To achieve better results, these methods extend and expand the features that CSP method can extract on EEG signals.

However, the above methods usually do not modify the CSP algorithm itself. Neither the shift in the frequency band or the time window, nor the addition of some regularization will change the purpose of the CSP algorithm, which is to distinguish the variance of the two classes of EEG signals by finding spatial filters. Therefore, to make up for these deficiencies, this paper introduces the local best particle swarm optimization algorithm to solve the spatial filters. These spatial filters can distinguish the general energy of the EEG signal in each frequency band, and are used as the supplement of the spatial filters obtained by the CSP algorithm [42]. In order to minimize the within-class matrix and maximize the between-class matrix of the energy of the EEG signal after spatial filtering, the adaptive spatial filters was calculated through continuous iteration. Based on FBCSP algorithm, a reasonable algorithm framework is designed by adding ASP features. The proposed FBCSP-ASP algorithm can not only select the subject-specific optimal spatial filter to improve the accuracy of MI-BCI classification, but also reduce the feature dimension for different subjects and suppress the negative impact of noise.

The contributions of this paper are listed as follow:

(1) A novel spatial filter solution framework is proposed for feature extraction of MI-EEG signals, which uses local optimal particle swarm optimization algorithm to solve the ASP spatial filter to distinguish the overall energy characteristics of different types of EEG signals.

(2) In order to improve the classification efficiency, the redundant features unrelated to MI are dropped. A two-stage feature selecting method based on MIBIF and DT-RFE is utilized for FBCSP-ASP features. So as to achieve faster, more accurate and more robust classification of MI tasks.

(3) In order to verify the feasibility and effectiveness of the proposed FBCSP-ASP algorithm, two public benchmark MI-EEG datasets of 9 subjects are selected for classification experiments, and the proposed algorithm shows accurate and robust results.

(4) With feature visualization, we analyze the differences and connections between the proposed ASP features and the traditional features, thus verify their complementarity.

The rest of the paper is organized as follows: Section 2 describes the methodology of the paper, including the extracted FBCSP-ASP features as well as the overall algorithm framework. Section 3 presents the results of the experiment and analyzes the results. Section 4 discusses the proposed method. Section 5 concludes the paper and points out the future work.

# 2 Methodology

## 2.1 Feature Extraction

We propose the FBCSP-ASP method as the feature extraction method, and for each frequency band, the features are extracted using CSP and ASP methods separately and merged. As a method that is proven effective on MI tasks, the FBCSP method can extract the energy difference between different leads for different types of MI tasks. On the other hand, the ASP method is used to extract the difference in the total energy of the leads for different types of MI tasks. We use the ASP algorithm as a feature complement to the FBCSP algorithm to improve the overall effect of the algorithm.

### 2.1.1 Common spatial pattern and filter bank common spatial pattern

CSP algorithm is a spatial filtering feature extraction algorithm for two-class classification tasks, which can extract the spatial distribution components of each class from multi-channel EEG signals. CSP algorithm designs a spatial filter to maximize the difference of variance values between two types of EEG signal matrices after spatial filtering, so as to obtain features with high discrimination. Detailed formulas of CSP is described in Appendix A. For multi-category MI tasks, one-vs-rest (OVR) strategy was used to extend the CSP algorithm [40]. FBCSP is an extension of CSP method, which executes CSP algorithms in different sub-bands to obtain FBCSP features. Therefore, for a k-class MI classification task with channels number of channels, FBCSP will obtain the features of subbands*k*channels in the preset subbands.

### 2.1.2 Local best particle swarm optimization

Particle swarm optimization (PSO) is an evolutionary computation technique. Compared with other optimization algorithms, PSO has no restrictions on the form and nature of the objective function and does not require gradient information [42]. It comes from the study of bird predation behavior. The basic idea of PSO is to find the optimal solution through the cooperation and information sharing among individuals in the swarm. PSO simulates a bird in a flock by designing a massless particle with only two attributes: speed and position. The speed represents the moving vector and the position represents solution. Each particle searches the optimal solution in the search space independently, which is recorded as the current individual best value. The individual best value is shared with other particles in the whole particle swarm, and the individual best value found is the current global optimal solution of the whole particle swarm. All particles in the PSO adjust their velocity and position according to the current individual extremum found by themselves and the current global optimal solution shared by the entire PSO.

PSO is initialized with a population of random particles as random solutions. The optimal solution is then found by iteration. In each iteration, the particle updates itself by keeping track of two extreme values: (*pbest, gbest*). After finding these two optimal values, the particle updates its velocity and position by the following formula:

$$v_i^{t+1} = \omega^t * v_i^t + c_1 \times rand(t) \times (pbest_i^t - x_i^t) + c_2 \times rand(t) \times (gbest_i^t - x_i^t) \quad (1)$$

$$x_i^{t+1} = x_i^t + \max(v_i^{t+1}, v_{max}) \quad (2)$$

$$\omega^t = \frac{(\omega_{ini} - \omega_{end})(G - t)}{G} + \omega_{end} \quad (3)$$

Eq. 1 represents the velocity-vector update formula, while Eq. 2 represents the position update formula at time t. Eq. 3 is the formula for calculating the inertia factor at time t using a linearly decreasing weight strategy (LDW).

In Eq. 1, $v_i^t$ is the original velocity vector of particle i at time t. $rand(t)$ is a random number between 0 and 1 used to increase the randomness of the algorithm. $\omega^t$ is the inertia factor, which represents the degree of dependence of the updated velocity vector on the original velocity vector. $x_i^t$ represents the current position of particle *i*, while $pbest_i^t$ and $gbest_i^t$ represent the personal best and global best positions of particle i at time t, respectively. $c_1$ and $c_2$ are the learning factors that represent the degree of learning of individual and global best values. In Eq. 2, the maximum displacement of particles is limited by $v_{max}$ during each iteration, and $x_i^{t+1}$ is updated iteratively. In Eq. 3, $G$ represents the maximum number of iterations, and $\omega_{ini}$ and $\omega_{end}$ represent the initial and final values of the inertia weight, respectively.

In this study, the Local Best PSO algorithm was employed, which differs from the traditional PSO algorithm in that it defines the global best value as the best value of the $k$-nearest particles around each particle, rather than the true global best value. This results in a longer convergence time but reduces the risk of the algorithm being trapped in local optima. Specifically, the definition of $gbest_i^t$ is as follows:

$$gbest_i^t = best(i, neighbor = k) \qquad (4)$$

Local Best PSO is summarized in **Algorithm 1**:

**Algorithm 1: Local Best PSO**

**Input: Particle number N**

**Output: Global best position gBest**

**Steps:**

(1) Initialize particles with random positions $x_i$ and velocities $v_i$ for each $i$ in range (N).
(2) Evaluate each particle and set the personal best position $pbest_i$ to $x_i$.
(3) Determine the best neighbor particle k and get $gbest_i$ for each particle $i$ by Eq. (4), using a topology like ring or star.
(4) Update $\omega$ using Eq. (3).
(5) Update particle's velocity $v_i$ and position $x_i$ using Eq. (1) and Eq. (2) respectively.
(6) While stopping criterion is not met, repeating step(2)-step(5)
(7) **Return** $gbest$

## 2.1.3 Adaptive spatial pattern

The CSP algorithm enhances the differences in variance values between two types of EEG signal matrices by designing a spatial filter and uses it to extract features. These features are in line with the requirements of the ERD/ERS phenomenon for decoding MI tasks, but they also pose some problems. For a binary classification problem of MI EEG signals, consider a data matrix of (*samples, channels, timepoints*). The CSP algorithm can only obtain a fixed (*channels, channels*) spatial filter, at most resulting in a (*channels, 1*) feature vector. Moreover, the objective of the spatial filter obtained by the CSP algorithm is only to distinguish the variance values of the EEG signal matrix. Therefore, the features extracted by CSP algorithm are very limited and not enough to decode MI-EEG signals well. To address this problem, we propose a new spatial-filter solving paradigm based on the PSO algorithm to complement the CSP spatial filter. We first establish a standard paradigm of EEG classification based on spatial filtering:

**Algorithm 2: EEG classification based on spatial filter**

**Input:** Raw training data, raw testing data
**Output:** Predicted testing label
**Steps:**
(1) Preprocess raw training data to obtain processed training data.
(2) Initialize Spatial Filters.
(3) While stopping criterion not met:
   a. Apply Spatial Filter to processed training data to obtain filtered training data.
   b. Compute the loss between filtered training data and training labels.
   c. Minimize the loss by updating the Spatial Filter.
(4) Save the learned Spatial Filters.
(5) Apply the learned Spatial Filters to the processed training data to obtain filtered training data.
(6) Extract features from the filtered training data to obtain train features.
(7) Train a classifier using the train features and training labels.
(8) Save the trained classifier.
(9) Preprocess raw testing data to obtain processed test data.
(10) Apply the learned Spatial Filters to the processed test data to obtain filtered test data.
(11) Extract features from the filtered test data to obtain test features.
(12) Use the trained classifier to predict the labels of the test features.
(13) **Return** Predicted testing label.

Both the CSP algorithm and the proposed ASP in this paper conform to the aforementioned standard paradigm for spatial filtering. For the CSP algorithm, the loss function is the difference in variance values between the two types of EEG signal matrices after filtering. For the ASP, we define the loss function as follows:

$$Loss = \frac{trace\left(\sum_{k=1}^{K}\sum_{i=1}^{n_k}(x_i - \bar{x}_k)(x_i - \bar{x}_k)^T\right)}{trace\left(\sum_{k=1}^{K} n_k(\bar{x}_k - \bar{x})(\bar{x}_k - \bar{x})^T\right)} \quad (5)$$

The numerator represents the within-class matrix of $K$-class MI signals after spatial filtering, while the denominator represents the between-class matrix of $K$-class MI signals after spatial filtering. The objective of the loss function is to make similar the signals of the same class of MI after spatial filtering while making different the signals of different classes of MI after spatial filtering. $x$ represents the feature extracted after spatial filtering, and in this paper, we use energy:

$$x = \log\left(\sum_{t=1}^{timepoints} |(F * X)|^2\right) \quad (6)$$

Where $F$ represents the spatial filter, and $X$ represents the EEG signals before spatial transformation. The process of obtaining the spatial filter in the ASP algorithm is conducted using **Algorithm 1**. In addition, considering the influence of frequency bands on MI signal energy, we use the same frequency band settings as FBCSP before applying the ASP algorithm, i.e., performing the ASP algorithm in each frequency band. Since one-vs-one (OVO) method can be performed for any two types of MI-EEG signals, more features and matrices can be obtained. In addition, redundant features will be removed by feature selecting, therefore, for the multi-class MI task, we adopt the one-vs-one (OVO) approach to implement the ASP algorithm. Thus, through the FBASP algorithm, we can obtain $subbands * C_k^2$ features, where $subbands$ is the number of frequency-bands, and $k$ is the number of task categories.

## 2.2 Feature Selection and Classifier

For our FBCSP-ASP method, we designed a two-stage feature selection strategy. As the number of FBCSP features that have not been selected is $subbands * k * channels$, which is much larger than the $subbands * C_k^2$ features obtained from FBASP. Moreover, FBCSP features have many redundant features [30-34]. Therefore, we first use a pre-set mutual information-based best individual feature (MIBIF) method to screen FBCSP features at each frequency band. The purpose of MIBIF is to retain effective FBCSP features while reducing the complexity of subsequent processing. Furthermore, after frequency band-level MIBIF is used to screen FBCSP features, the decision tree-based recursive feature elimination (DT-RFE) method is used for the second-stage feature selection of all sub-band FBCSP-ASP features. DT-RFE is used to select effective features that are suitable for the subject, to choose spatial filters that are more helpful for classification tasks.

The time complexity of the second stage DT-RFE method is $O(n * features^2)$, where $n$ is the sample size and $features$ is the number of features. While not using MIBIF, $features$ is the sum of the number of features of FBCSP and FBASP which is $subbands * k * channels + subbands * C_k^2$, MIBIF greatly reduces the time complexity of DT-RFE by reducing the dimension of FBCSP to the same order of magnitude as FBASP and reducing number of features from $subbands * k * channels + subbands * C_k^2$ to $constant * subbands * C_k^2$.

## 2.2.1 Mutual information-based best individual feature

To reduce the dimensionality of FBCSP features, we used mutual information-based best individual feature (MIBIF) as the feature selection method in each frequency band. MIBIF is a feature selection method based on mutual information. In MIBIF, the $n$ features with the highest mutual information are selected from the feature vectors obtained from the $k$ projection matrices in each frequency band. Mutual information is calculated as follows:

$$I(X;Y) = \sum_{y \in Y} \sum_{x \in X} p(x,y) \log\left(\frac{p(x,y)}{p(x)p(y)}\right) \qquad (7)$$

Here, X and Y are the features and corresponding labels obtained from each OVR projection matrix. After going through MIBIF, FBCSP features can obtain $subbands * k * n$ FBCSP features, where $subbands$ is the number of frequency bands, $k$ is the number of task categories, and $n$ is the number of selected features in each projection matrix.

## 2.2.2 Recursive feature elimination

We used the Recursive Feature Elimination (RFE) method to select spatial filters that would be more helpful for the classification task. FBCSP and FBASP each generated $N_{FBCSP}$ and $N_{FBASP}$ spatial filters, and we used RFE to select $N_{FBCSP-ASP}$ better spatial filters from them. RFE is a machine learning feature selection algorithm used to build models and reduce computation time, coefficient number, and model complexity. It is an improvement technique for filter methods, especially for feature correlation coefficient screening and filter methods based on L1 regularization. It uses an internal algorithm to recursively eliminate unimportant features. In RFE, at each iteration, a model based on the current best feature subset is constructed. Then, in each iteration, the model sorts each feature according to its importance. Higher-ranked features are retained, and lower-ranked features are recursively removed. The process of collecting important features and iterative model improvement results in the final optimal feature subset. The algorithm implementation of RFE is as follows:

**Algorithm 3: Recursive feature elimination**
**Input:** FBCSP features, FBASP features, $N_{FBCSP-ASP}$
**Output:** Optimal set of features
**Steps:**
(1) Initialize feature set with all $N_{FBCSP}+N_{FBASP}$ features.
(2) Train the model with the current feature set.
(3) Compute the importance of each feature in the trained model.
(4) Drop the feature with the lowest importance score from the feature set.
(5) Repeat step (2)-step (4) for $(N_{FBCSP}+N_{FBASP}-N_{FBCSP-ASP})$ times.
(6) **Return** the optimal set of features with the desired number of features.

## 2.2.3 Decision tree based recursive feature elimination and random forest

We chose Decision tree (DT) as the internal model for RFE, and we used random forest (RF) as the classifier for the model. We selected tree-based models for both feature selection and classification because the features extracted from FBCSP and FBASP are of different orders of magnitude, and tree-based models process features vertically and are not affected by differences in feature magnitude. On the other hand, the loss function used for ASP features is based on the between-class matrix and the within-class matrix, making the tree-based model based on the node value suitable for FBCSP-ASP features.

DT is a machine learning classification method based on a tree structure. In DT, classification is performed by iterative splitting of data. Each node from the root node to the leaf node represents a split. For DT, it is necessary to keep the data with the same class as much as possible on one side of the tree. When the data in the leaf node of the tree are all of the same class, classification stops. In this paper, the splitting of DT nodes is based on the Gini coefficient:

$$Gini(p) = \sum_{k=1}^{k} p_k(1-p_x) = 1 - \sum_{k=1}^{k} p_x^2 \qquad (8)$$

Here, $k$ represents the number of classes. $p_k$ represents the probability of a particular class in the current category, and $1-p_x$ represents the probability that it is not the current class. The larger the Gini coefficient value, the greater the uncertainty of the sample. By calculating the Gini coefficient, we select the attribute that minimizes the Gini coefficient after splitting as the optimal splitting point. Meanwhile, the feature importance in RFE is obtained by calculating the normalized decrease in the Gini coefficient for each feature. For the features used in splitting each node in the decision tree, their feature importance is calculated as follows:

$$Importance = \frac{N_t}{N} * \left(Gini - \frac{N_{tL}}{N_t} * left_{Gini} - \frac{N_{tR}}{N_t} * right_{Gini}\right) \qquad (9)$$

Where $N$ represents the number of samples, $N_t$ represents the number of samples in the current node. $Gini$ represents the Gini coefficient of the current node. $N_{tL}$ represents the number of samples in the left child node of the current node. $left\_Gini$ represents the Gini coefficient of the left child node of the current node. $N_{tR}$ represents the number of samples in the right child node of the current node, and $right\_Gini$ represents the Gini coefficient of the right child node of the current node.

Random Forest (RF) is an ensemble learning model based on bagging, with Decision Tree (DT) as the base classifier. The process of generating decision trees in random forest involves both row and column sampling of the sample data. By randomly selecting a part of the dataset, a tree is generated, and repeating this process generates different decision trees, which together form the random forest. In the output, the final output of the RF is the collective decision results of the decision trees obtained by voting. The training process of the random forest is as follows:

**Algorithm 4: Random forest**

**Input: Dataset N, Number of decision trees T, Number of randomly sampled features F**

**Output: Trained model RF**

**Steps:**

(1) Initialize an empty list for decision trees, DTs.

(2) Randomly sample F features from N to create a new dataset N'.

(3) Create a new decision tree DT using dataset N'.

(4) Append DT to DTs.

(5) Repeat step (2)-step (4) for T times.

(6) Create the random forest classifier RF by uniformly selecting from the decision trees in DTs.

(7) **Return** trained RF.

## 2.3 FBCSP-ASP

In summary, we integrate the feature extraction, feature selection, and classifier discussed above. Continuing previous research [40]. Same as settings in baseline, 9 subbands were set, ranging from 4–8 Hz, 8–12 Hz to 36–40 Hz. The training and testing framework of the FBCSP-ASP algorithm is shown in the figure below:

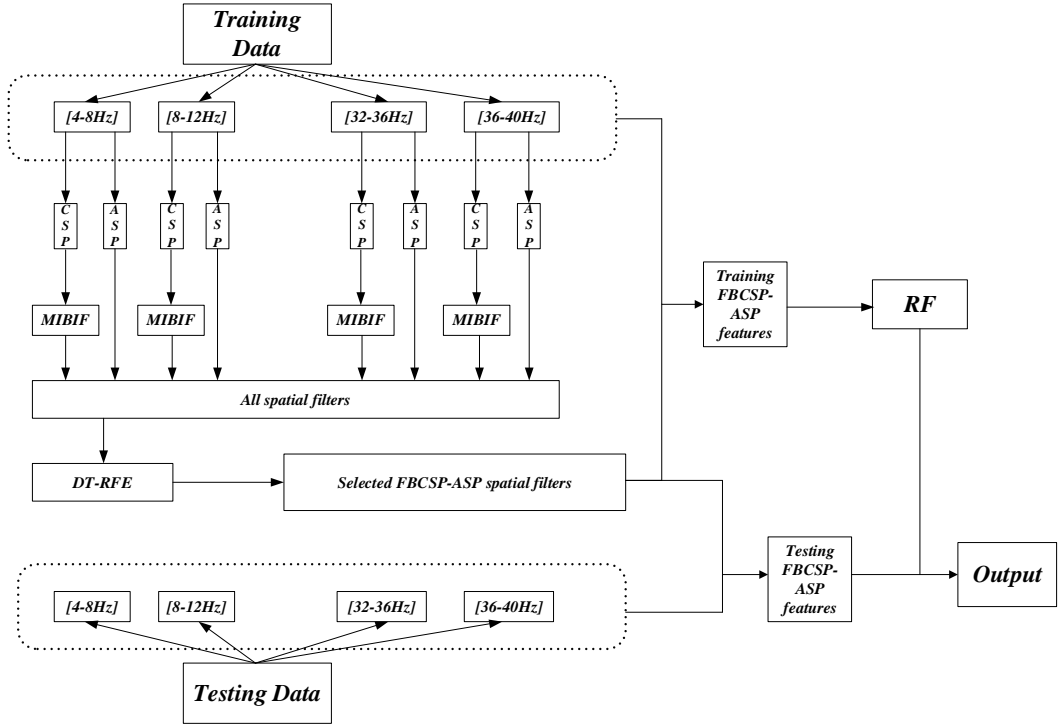

**Fig. 1.** The overall framework for the proposed FBCSP-ASP algorithm.

During the training stage, the original signal used for training is first filtered into 9 sub-bands. CSP and ASP features are then separately computed on each sub-band. The CSP features on each sub-band are subjected to the first round of feature selection using the MIBIF method to coarsely reduce the total feature dimension. This can greatly reduce the computation of the subsequent DT-RFE method while retaining effective CSP features. The ASP spatial filters and the selected CSP spatial filters on each sub-band are merged. The merged CSP and ASP features on each sub-band are then pooled together, resulting in a total of $subbands * (C_k^2 + k * n)$ feature vectors. These feature vectors are sent to DT-RFE for dimension reduction, and the optimal number of features after dimension reduction is determined through 5-fold validation on the training set. Then, the classifier is trained on these features and saved. During the testing stage, the original signal used for testing is filtered into the same sub-bands, and the saved FBCSP-ASP spatial matrices are used to extract the corresponding features in each sub-band. These features are classified by the trained classifier to obtain the final output.

# 3 Experiments and results

## 3.1 Datasets

In order to verify the feasibility and effectiveness of our proposed FBCSP-ASP algorithm, we conducted experiments on two publicly available benchmark datasets from BCI Competition IV: dataset 2a and dataset 2b [43]. Both datasets contain MI-EEG signals collected from 9 different subjects. Dataset 2a contains 4 MI tasks (left hand, right hand, feet, and tongue), while dataset 2b contains 2 MI tasks (left hand and right hand). Each subject in dataset 2a has 2 sessions, and each session has 72 trials across four categories. Each subject in dataset 2b had 5 sessions, with 2 sessions containing 120 trials and the remaining 3 sessions containing 160 trials. The experimental paradigms for performing MI tasks in both datasets are shown in Fig. 2. Initially, a warning beep and a cross were presented on the screen to keep the subjects focused. Then, an arrow prompt was displayed on the screen to guide the classification of the MI task. After a 1-second prompt, the subjects began to perform the MI task according to the guidance, which lasted for 4 seconds. Upon completing the MI task, the subjects entered a rest period. It should be noted that the last three sessions of dataset 2b included smiley feedback, but this feature was not specifically addressed in this paper.

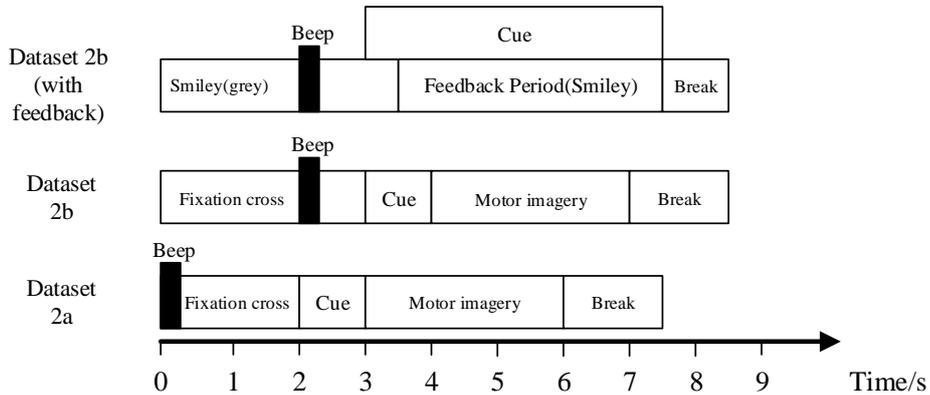

**Fig. 2.** Experimental paradigm.

## 3.2 Experiments Setups

For both datasets, we selected the EEG signals from 0.5 s to 3.5 s after the cue as the input for the algorithm. In dataset 2a, for any given subject, we used the first session as the training set and the second session as the testing set. In dataset 2b, for any given subject, we used the first three sessions as the training set and the last two sessions as the testing set. The two datasets are described in Table 1.

**Table 1**. Description of datasets.

| | Dataset 2a | | | Dataset 2b | | |
|---|---|---|---|---|---|---|
| Training set | | | Testing set | Training set | | Testing set |
| 288 | | | 288 | 400 | | 320 |
| Electrodes | Sampling rate | Duration | | Electrodes | Sampling rate | Duration |
| 22 | 250 Hz | 3000 ms | | 3 | 250 Hz | 3000 ms |
| Frequency bands | | | | Frequency bands | | |
| 4–8 Hz, 8–12 Hz, …, 36–40 Hz | | | | 4–8 Hz, 8–12 Hz, …, 36–40 Hz | | |

The hyperparameter settings required for the experiment involve the PSO algorithm, the MI-based dimensionality reduction after FBCSP, the RFE algorithm, and the RF classifier. Among them, some hyperparameters are obtained by the mesh parameter tuning method with 10-fold validation on the training set. The hyperparameter settings is described in Appendix B.1. Among them, the PSO parameter settings are obtained according to the conclusion of research [44], so as to achieve the global optimum faster and avoid the local optimum as much as possible.

## 3.3 Comparison results of motor imagery classification

In order to compare the classification performance and efficiency of our proposed algorithm, we used FBCSP+SVM algorithm and FBASP+RF as baseline1 and baseline2, respectively. Additionally, we compared our proposed algorithm with Deep ConvNet [45], Shallow ConvNet [45], EEGNet [46], C2CM [47], and STNN [48]. The results on dataset 2a are presented in the table below:

**Table 2.** Results on dataset 2a.

| | Baseline1 | Baseline2 | Deep ConvNet | Shallow ConvNet | EEGNet | C2CM | STNN | Proposed |
|---|---|---|---|---|---|---|---|---|
| A1 | 77.4 | 82.3 | 86.6 | 79.5 | 85.0 | **87.5** | 82.3 | **87.5** |
| A2 | 54.2 | 47.6 | 62.3 | 56.3 | 56.6 | **65.3** | 47.6 | 59.0 |
| A3 | 69.8 | 84.3 | 89.9 | 88.9 | 81.7 | 90.3 | 88.9 | **90.6** |
| A4 | 56.3 | 66.6 | 65.6 | **80.9** | 66.4 | 66.7 | 60.8 | 67.4 |
| A5 | 46.9 | 55.5 | 55.2 | 57.3 | 54.9 | 62.5 | **66.7** | 63.2 |
| A6 | 52.1 | 49.6 | 48.5 | 53.8 | 59.6 | 45.5 | **57.9** | 57.3 |
| A7 | 83.0 | 62.5 | 86.1 | 91.7 | **92.3** | 89.6 | 85.8 | 83.3 |
| A8 | 60.4 | 75.7 | 78.4 | 81.2 | 75.7 | **83.3** | 77.1 | 80.2 |
| A9 | 68.4 | 72.1 | 76.1 | 79.2 | 74.8 | 79.5 | 80.9 | **83.0** |
| Mean | 63.17 | 66.24 | 72.10 | 74.31 | 71.89 | 74.47 | 72.20 | **74.61** |
| SD | 12.19 | 12.74 | 14.83 | 14.54 | 13.28 | 15.33 | 13.48 | 12.13 |

The results on dataset 2b are presented in the table below:

**Table 3.** Results on dataset 2b.

| | Baseline | Baseline2 | Deep ConvNet | Shallow ConvNet | EEGNet | C2CM | STNN | Proposed |
|---|---|---|---|---|---|---|---|---|
| B1 | 70.3 | 69.4 | 72.0 | 74.2 | 73.8 | **74.8** | 85.0 | 74.4 |
| B2 | 55.4 | 58.9 | 57.0 | 55.8 | 56.7 | 61.3 | 75.2 | **62.5** |
| B3 | 55.6 | 60.9 | 64.9 | 55.4 | 64.5 | 65.5 | **68.2** | 61.2 |
| B4 | 94.7 | 88.5 | 94.4 | 91.6 | 93.2 | 94.4 | **98.9** | 97.1 |
| B5 | 80.6 | 85.8 | 89.9 | 88.7 | 81.9 | 86.7 | 75.0 | **90.3** |
| B6 | 80.0 | 83.0 | 83.3 | 83.3 | 85.8 | **87.5** | 82.0 | 85.0 |
| B7 | 74.1 | 76.5 | 78.1 | 74.1 | 72.7 | 79.4 | 83.2 | **83.8** |
| B8 | 79.7 | 86.1 | 90.8 | 88.6 | **91.5** | 89.6 | 79.5 | 88.6 |
| B9 | 76.3 | 79.2 | 77.9 | 72.8 | 72.5 | 81.7 | 79.0 | **87.8** |
| Mean | 74.08 | 76.47 | 78.70 | 76.09 | 76.96 | 80.10 | 80.7 | **81.19** |
| SD | 12.47 | 10.41 | 12.48 | 11.95 | 12.20 | 11.13 | 8.50 | 11.79 |

From the results in Tables 2 and 3, it can be seen that the proposed FBCSP-ASP method has achieved satisfactory performance on datasets 2a and 2b, reaching accuracies of 74.6% and 81.2%. The classification performance of the proposed algorithm is significantly improved than baselines using a statistical student t-test ($p < 0.05$) on both datasets. Interestingly, when comparing the performance of the two baselines on the two datasets, we found that some subjects were particularly suited to either CSP or ASP features. For example, in dataset 2a, subject a3 achieved significantly better results with FBASP (84.3%) than with FBCSP (69.8%), while subject a7 showed the opposite trend, with FBCSP (83.0%) outperforming FBASP (62.5%). Compared to the two baselines, the proposed FBCSP-ASP algorithm achieved better classification performance by not only adding additional spatial filters, but also by removing redundant features using the DT-RFE algorithm. Moreover, on average, we found that the proposed FBCSP-ASP algorithm outperforms traditional deep learning models such as EEGNet. Additionally, it shows results that are comparable to those of recent models such as EEGNet, C2CM, and STNN. Besides, since recent deep learning models have introduced more complex network architectures, these models have more complex in interpretability.

## 3.4 Results analysis

We analyze the results from three perspective. Firstly, we conducted an analysis of the FBCSP-ASP features before performing DT-RFE feature selection at the individual level. We explored the performance of the ASP and CSP features in different frequency bands using a three-dimensional histogram of mutual information. Next, we visualized the features to investigate the effectiveness of the FBCSP-ASP algorithm. Finally, we investigated the contribution of the FBCSP-ASP features to classification at the model level. Subject A3 and subject B5 were selected as representative for dataset 2a and 2b for detailed analysis.

3.4.1 Feature-level analysis by mutual information

To investigate the performance of ASP and CSP features in different frequency bands, as well as to demonstrate the effectiveness of the ASP algorithm, we analyzed the FBCSP-ASP features of each subject before DT-RFE feature selection using a mutual information-based approach. Mutual information was calculated using Eq. 7. For each subject in dataset 2a and dataset 2b, we calculated the mutual information of each feature and plotted them in the form of a three-dimensional bar graph.

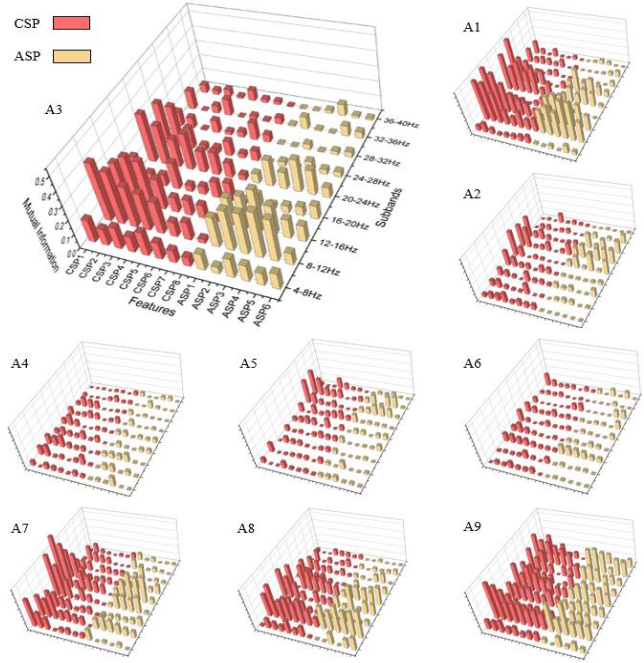

**Fig. 3.** Mutual information 3D plot for dataset 2a. CSP features are colored red and ASP features are colored yellow. The position of the cylinder on the plane represents the type and frequency band of the feature, and the height of the cylinder represents the mutual information value of the feature. The subject number represented by each subfigure is marked at the top left of the subfigure.

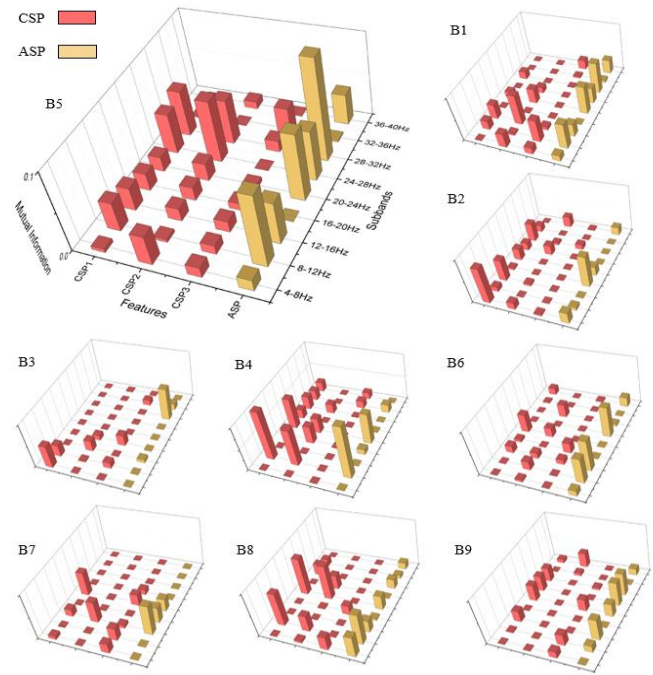

**Fig. 4.** Mutual information 3D plot for dataset 2b.

Fig. 3 and Fig. 4 present the mutual information of all FBCSP and FBASP features for each subject in datasets 2a and 2b. The CSP and ASP features are marked in red and yellow. We use A3 and B5 as examples for the four-class and binary-class analysis. For A3, the mutual information values in the frequency bands of 8–12 Hz, 12–16 Hz and 20–24 Hz are higher than others are. For subject A3, the mutual information values of CSP features are slightly higher than those of ASP features in all frequency bands. This indicate that CSP features perform better on subject A3. However, for subject B5, ASP features have higher mutual information values than CSP features in the 8-16Hz and 20-23Hz frequency bands. Considering that CSP features extract differential features through spatial transformation, while ASP features extract overall energy features through spatial transformation, this suggests that the suitable types of features for subjects during MI tasks is different.

In summary, the sub-bands where different subjects' features perform well are not the same, but FBCSP and ASP features generally have similar changes in mutual information at the frequency band level. On the other hand, although FBCSP has higher mutual information values than ASP in most sub-bands, ASP still has equally high mutual information values. Even in subjects in dataset 2b, the ASP features in the same sub-band may have higher mutual information values.

3.4.2 Feature visualization

To explore the optimization effects of the FBCSP-ASP algorithm on features, we utilized the dimensionality reduction visualization tool t-SNE [49] for feature visualization. Fig. 5 and Fig. 6 show the visualization of different classes of features under the optimal FBCSP-ASP features for both datasets. For horizontal comparison, we also present the results of the classic FBCSP features and the use of FBASP features only. In terms of visualization, t-SNE tool was used to reduce the features of each group of EEG signals to two dimensions, and displayed in scatter plots.

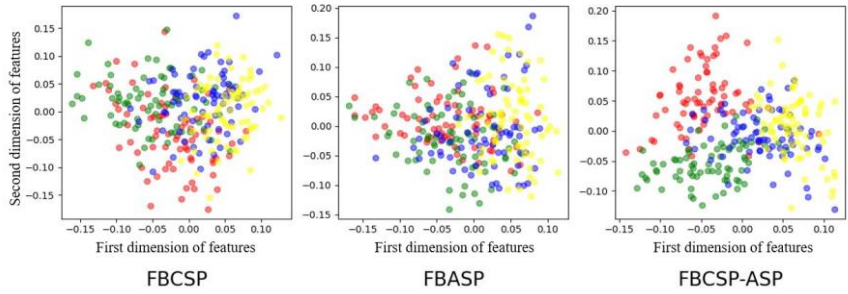

**Fig. 5.** Feature visualization on dataset 2a. The colors representing the left hand, right hand, foot and tongue are red, green, blue, and yellow.

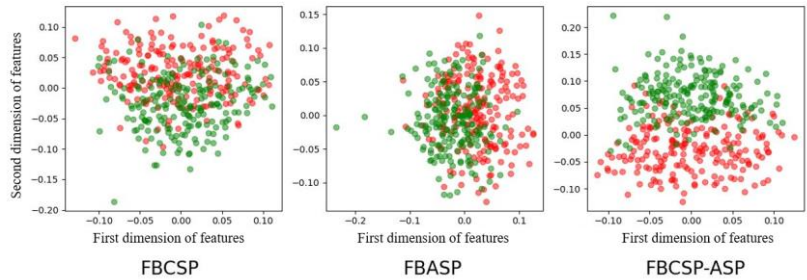

**Fig. 6.** Feature visualization on dataset 2b. The colors representing the left hand and right hand are red and green.

The results from Fig. 5 and Fig. 6 demonstrate that the proposed FBCSP-ASP features outperform traditional FBCSP features, for datasets 2a and 2b. The selected FBCSP-ASP features exhibit better separability in t-SNE visualization both for the 4-classes in dataset 2a and for 2-classes in dataset 2b. It means that the discriminability of the FBCSP features, which serve as the baseline, is weaker than the selected FBCSP-ASP features. On the other hand, neither using only FBCSP nor only FBASP features can fully distinguish between classes. However, after combining FBCSP-ASP features, the features are better clustered by class, which demonstrates the complementarity of CSP and ASP features.

3.4.3 Feature contribution analysis by Shapley values

In order to prove the improvement of FBCSP by the addition of ASP, we use Shapley values to explore the contribution impact of each feature on the final classifier. Shapley additive explanations (SHAP) is a model explanation method from game theory [50], which is proven to achieve interpretability for machine learning [51]. Considering a situation where a coalition of players co-create value and reap benefits, SHAP gives a calculation method to distribute the benefits. SHAP allocates expenditure to players according to their contribution to the total expenditure. For a regression model, all input variables contribute to the final prediction, so every variable is a player in the coalition. The prediction is the co-create value of coalition. The importance of variables, which namely SHAP values are measured by how much it contributes to the prediction. For sample $x$, SHAP value of variable $j$ is calculated by following formula:

$$\Phi_j(val) = \sum_{S \subseteq \{x_1,\dots,x_p\}\{x_j\}} \frac{|S|!(p-|S|-1)!}{p!} \left(val(S \cup \{x_j\}) - val(S)\right) \quad (10)$$

Where $val$ is a specific model and $\Phi_j(val)$ is the SHAP values of variable j with this model. $S$ is a subset of input variables. $|S|$ is the number of variables in subset $S$. $p$ is the total amount of variables in the prediction model. The global SHAP value of variable $j$ is the sum of absolute SHAP values of $j$ among all samples. Therefore, for an $n$ classification task, we will get SHAP values for each of the $n$ classes.

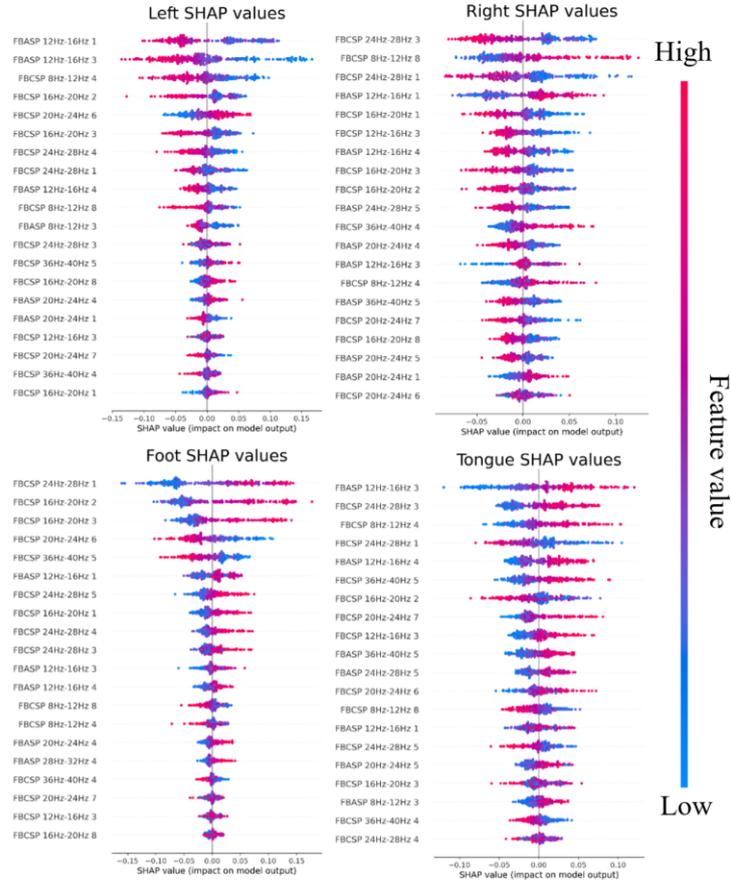

**Fig. 7.** SHAP values for FBCSP-ASP features on dataset 2a. Each subplot represents the contribution of features to the classification of a type of MI task. For each subplot, ranking of features

is from top to bottom according to the mean absolute value of SHAP in all samples.

Fig. 7 illustrates the SHAP values of the FBCSP-ASP method for MI four-class classification. In any row of the subplot, each point represents a sample, and its lateral position is the SHAP value calculated for the corresponding feature on that sample. The color of the point represents the ranking of the corresponding feature value in all samples, ranging from blue-purple-red. It can be seen that for distinguishing right hand and foot, CSP-based features have a greater advantage. However, on the other hand, for distinguishing left hand and tongue contribution to the head, the introduction of ASP features contributes more to the classifier. Taking the SHAP values for distinguishing left hand in Fig. 7 as an example, the "FBASP 12Hz-16Hz 1" feature has an obvious break at 0, indicating that this feature has almost no redundant information for the samples. At the same time, when the feature calculated by "FBASP 12Hz-16Hz 1" is higher, it tends to be classified as non-left-hand motor imagination, and when it is lower, it tends to be classified as left-hand motor imagination. Compared with the best FBCSP-based feature in the same plot, "FBCSP 8Hz-12Hz 4", which also has good separability, cannot provide valuable information to the classifier for samples with feature values in the center. Considering the calculation process of CSP and ASP algorithms, it can be inferred that for the left-hand discrimination task, the spatial filters obtained by the ASP algorithm have better performance than those obtained by the CSP algorithm.

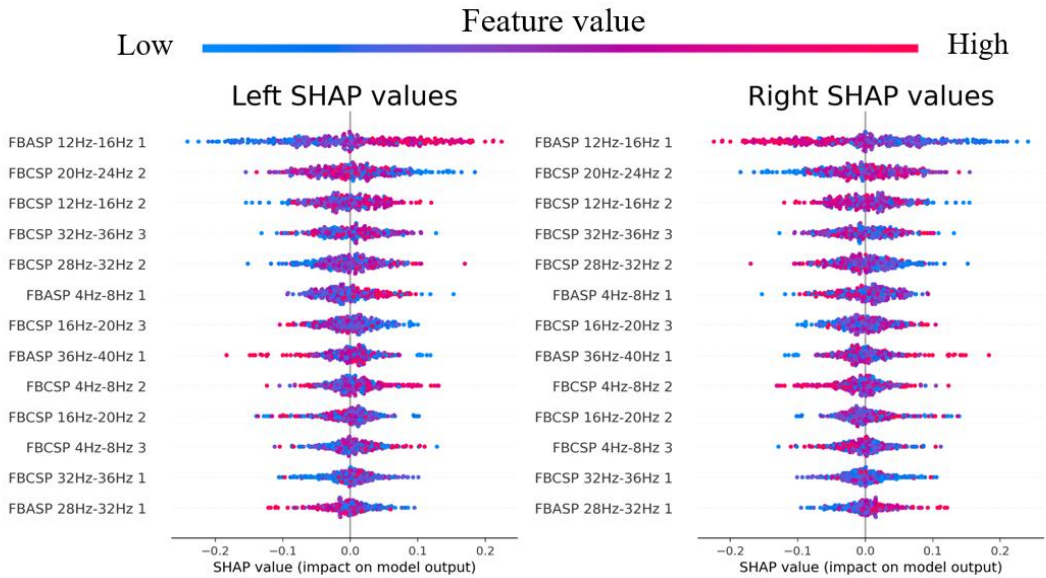

**Fig. 8.** SHAP values for FBCSP-ASP features on dataset 2b.

Fig. 8 illustrates the SHAP values of the FBCSP-ASP method for binary classification in dataset 2b. For a binary classification task, the discriminative contribution of any feature to both classes is symmetric. As shown in Fig. 8, the ASP feature in the 12–16 Hz frequency band has the greatest effect on the classifier. On the other hand, apart from the ASP feature in the 12–16 Hz frequency band, most of the selected features are CSP features. From Fig. 8, it can be inferred that after the transformation by the corresponding ASP spatial filters, the energy of 12–16 Hz frequency band of EEG signal is larger during left hand motor imagery than that during right hand motor imagery. The SHAP values of binary classification indicate that although FBASP can produce better features, FBCSP features still have a significant effect.

# 4 Discussion

This paper proposes a novel spatial filter-based EEG signal feature extraction method, called the ASP method, and designs an FBCSP-ASP method for the classification of MI EEG signals. The proposed algorithm outperforms traditional machine learning based algorithms in the classification of MI and achieves excellent results on two datasets.

From the results of Fig. 6 and Fig. 7, we find that the features selected by the FBCSP-ASP algorithm show better classification performance on datasets 2a and 2b. Although the training process of FBCSP-ASP is more complex, the spatial filters selected by the FBCSP-ASP method can directly act on the filtered EEG signals during the testing phase. If the trained classification model is used for online applications, for each sample, spatial filtering can be performed on different frequency bands and energy features can be extracted for direct use in MI classification, resulting in higher classification performance.

In order to further investigate the relationship between ASP features and CSP features, and demonstrate that the features selected by FBCSP-ASP are the most discriminative. We employed 3D mutual information plot, t-SNE, and SHAP values to analyze and visualize the features. The 3D mutual information plot was used to visualize the relationship between ASP and CSP features at the feature level. T-SNE was used to analyze the differences in FBCSP/FBASP/FBCSP-ASP features from an intuitive perspective and transform them into 2D space. SHAP values were used to analyze the contribution of FBCSP-ASP features to the model. In the experimental results, we displayed the results of three different methods: FBCSP, FBASP, and FBCSP-ASP. The 3D mutual information plot calculates the mutual information between CSP and ASP features on each frequency band and the labels, which can intuitively display the relationship between ASP and CSP features. In t-SNE analysis, we intuitively found that FBCSP-ASP features improve upon both FBCSP and FBASP features, which validates the effectiveness of the FBCSP-ASP method and the complementarity between ASP and CSP features. SHAP values demonstrate the contribution of ASP and CSP features in different frequency bands to each class classification by calculating the contribution value of each feature to Eq. 10, which further validates the effectiveness of ASP features in the classifier.

The proposed FBCSP-ASP algorithm has three advantages:

(1) **Extendibility.** Our **Algorithm 1** breaks the tradition that only CSP algorithm can be used as a spatial filter and proposes a customizable process for calculating brain signal spatial filtering features. The ASP feature proposed in this paper is an instance of **Algorithm 2**, which uses **Algorithm 1** as a spatial filtering calculation method and Eq. 5 and Eq. 6 as loss functions. For further research, algorithms for finding spatial filters and loss functions for spatial filtering can be modified. Therefore, the ASP feature is very flexible and has strong expandability for different EEG signal classification tasks.

(2) **Generalization.** The FBCSP-ASP algorithm has certain generalization ability for MI-EEG signal classification. Compared with the baseline algorithm, our proposed algorithm has greatly improved average classification accuracy for all subjects on two datasets. In addition, although the training process of FBCSP-ASP algorithm is complex, once the CSP and ASP spatial filters on each frequency band are determined, they can be applied to the EEG signals of the entire subject collected on a certain acquisition device, showing good practicality in MI classification. Unlike deep learning models that require a large number of training experiments, the proposed FBCSP-ASP algorithm can be trained and applied in a small number of EEG experiments. Therefore, it has a broad application prospect and potential in wearable EEG devices, wireless transmission EEG devices, and many other application scenarios.
(3) **Interpretability.** The proposed FBCSP-ASP algorithm uses RF as a classifier, which enables us to analyze more detailed mutual information processes. Meanwhile, interpretable methods, such as SHAP values applicable to machine learning, can also be used for analysis. Such analysis can reveal which type of features have better performance on each subject and the contribution of different features to classification for each subject, as shown in Fig. 7 and Fig. 8. This interpretability is of great value to the research in the field of MI-BCI.

# 5 Conclusion

This study proposes the FBCSP-ASP method by utilizing the ASP spatial filter to obtain more valuable spatial features in different frequency bands, and combining them with traditional CSP features. The Local Best PSO algorithm is employed to optimize the designed loss function, to solve for spatial filters applicable for MI classification beyond CSP. To avoid redundant features and improve recognition efficiency, the CSP features of each frequency band are first preliminarily screened based on mutual information, and then merged with ASP features for further feature selection using the DT-RFE method.

The proposed FBCSP-ASP method is evaluated on two publicly available EEG datasets. The selected FBCSP-ASP features are superior to FBCSP features, and perform comparably to SOTA methods. At the same time, analysis based on mutual information, t-SNE, and SHAP are utilized to specifically analyze the response of each feature on the subjects. In the future, the proposed algorithm will be applied to online MI-BCI for stroke therapy.

While the proposed algorithm can improve the classification accuracy of the baseline, there is still room for improvement in the methods for solving spatial filters and the loss function chosen. Therefore, future work will include further research into variants of the ASP method, exploring better methods for solving spatial filters based on the proposed spatial filter solution paradigm, and investigating more effective loss functions.

# Data availability

The authors do not have permission to share data.

# Acknowledgements

We acknowledge financial support from Fundamental Research Funds for the Central Universities (2020XD-A06-1), State Key Program of the National Natural Science Foundation of China (82030037), National Natural Science Foundation of China (62203063), China Postdoctoral Science Foundation (2022M710464), and BUPT Interdisciplinary Team Cultivation Program.

# Appendix A

For a two-class EEG classification task, the two categories are denoted as matrices X1 and X2 respectively, and their shapes are *channels*time-samples*. The mathematical model of composite source is used to describe the EEG signal. Under the condition of ignoring the influence of noise, the two matrices can be described as follows:

$$X_1 = [C_1 \ C_M] \begin{bmatrix} S_1 \\ S_M \end{bmatrix} \quad (1)$$

$$X_2 = [C_2 \ C_M] \begin{bmatrix} S_2 \\ S_M \end{bmatrix} \quad (2)$$

Here, $S_1$ and $S_2$ represent the source signals that are linearly independent of each other in the two types of EEG signals, and $S_M$ is the source signal common to the two tasks. $C_1$ and $C_2$ are composed of the common spatial patterns corresponding to the respective sources of $S_1$ and $S_2$, respectively. Each spatial pattern is a vector of $channels * 1$, and the physical meaning of this vector represents the distribution weight of the signal caused by a single source signal on all channels. $C_M$ then represents the source signal common to the two types corresponding to $S_M$. The purpose of CSP is to find an optimal set of spatial filters for projection that maximizes the difference of variance values between two types of signals. The sum R of the normalized average covariance matrices of $X_1$ and $X_2$ is denoted as follows:

$$R = \overline{R_1} + \overline{R_2} = \frac{X_1 X_1^T}{trace(X_1 X_1^T)} + \frac{X_2 X_2^T}{trace(X_2 X_2^T)} \quad (3)$$

Where $X_1^T$ and $X_2^T$ are the transpose of $X_1$ and $X_2$, respectively, $trace$ denotes the trace of the matrix, with $\overline{R_1}$ and $\overline{R_2}$ denote the average covariance matrices of the covariance matrices $R_1$ and $R_2$ in the respective class experiments, respectively. The mixture spatial covariance matrix is obtained by eigenvalue decomposition:

$$R = U\lambda U^T \quad (4)$$

Where $U$ is the matrix of eigenvectors and $\lambda$ is the diagonal matrix formed by the corresponding eigenvalues. The eigenvalues are sorted in descending order, and the whitening value matrix $P$ is:

$$P = \sqrt{\lambda^{-1}} U^T \quad (5)$$

The covariance matrices $R_1$ and $R_2$ are transformed as follows:

$$S_1 = P R_1 P^T \quad (6)$$
$$S_2 = P R_2 P^T \quad (7)$$

After that, $S_1$ and $S_2$ are decomposed by principal components:

$$S_1 = B_1 \lambda_1 B_1^T \quad (8)$$
$$S_2 = B_2 \lambda_2 B_2^T \quad (9)$$

Here, $B_1$ and $B_2$ are the eigenvector matrices of $S_1$ and $S_2$, $\lambda_1$ and $\lambda_2$ are the corresponding eigenvectors of $B_1$ and $B_2$, respectively, and the sum of $\lambda_1$ and $\lambda_2$ is the identity matrix. That is, the sum of eigenvalues of the two types of matrices is always 1. Therefore, the covariance matrix of each trial data is whitened and projected along its whitened total covariance matrix to obtain the optimal projection transformation matrix, that is, the spatial filter $W$ is:

$$W = B^T P \quad (10)$$

For two types of data $X_1$ and $X_2$, the corresponding CSP feature vector is represented by:

$$\begin{cases} Z_1 = W * X_1 \\ f_1 = \dfrac{var(Z_1)}{sum(var(Z_1))} \end{cases} \quad (11)$$

$$\begin{cases} Z_2 = W * X_2 \\ f_2 = \dfrac{var(Z_2)}{sum(var(Z_2))} \end{cases} \quad (12)$$

Therefore, after CSP transformation, the spatial domain feature vector of $channels * 1$ can be obtained.

# Appendix B

**Table B.1** Description of hyperparameters.

|  | Hyperparameters | Range or value for Dataset 2a | Range or value for Dataset 2b |
|---|---|---|---|
| Local PSO | $c_1$ | 2 | 2 |
|  | $c_2$ | 2 | 2 |
|  | $w$ | 0.729 | 0.729 |
|  | $k$ | 20 | 15 |
|  | Particle lower bound | -100 | -100 |
|  | Particle upper bound | 100 | 100 |
|  | Number of particles | 100 | 60 |
|  | Iterations | 2000 | 2000 |
| MIBIF | Number of selected features for each subband | 8 | None |
| DT-RFE | Max depth | None | None |
|  | Number of selected features | 5 to 125 | 1 to 36 |
| RF | Number of DTs | 5 to 200 | 5 to 200 |
|  | Max depth | 1 to 20, None | 1 to 20, None |
|  | Min samples split | 1 to 10 | 1 to 10 |
|  | Min samples leaf | 1 to 5 | 1 to 5 |